\documentclass[aps,prl,amsmath,amssymb,reprint,superscriptaddress,floatfix]{revtex4-2}

\usepackage[utf8]{inputenc}
\usepackage{graphicx}
\usepackage{dcolumn}
\usepackage{bm}
\usepackage[hidelinks]{hyperref}
\usepackage{newtxtext,newtxmath}

\newcommand{\dd}{\mathrm d}

\newcommand{\Yeps}{Y_{\varepsilon'}}

\begin{document}

\title{Marginal Operator Deformations and Pseudocriticality in Baxter--Wu Models}

\author{Hyun-Yong Lee}
\email{hyunyong@korea.ac.kr}
\affiliation{Department of Applied Physics, Graduate School, Korea University, Sejong 30019, South Korea}
\affiliation{Center for Extreme Quantum Matter and Functionality, Sungkyunkwan University, Suwon 16419, South Korea}
\affiliation{Institute for Solid State Physics, The University of Tokyo, Kashiwa, Chiba 277-8581, Japan}

\author{Naoki Kawashima}
\email{kawashima@issp.u-tokyo.ac.jp}
\affiliation{Institute for Solid State Physics, The University of Tokyo, Kashiwa, Chiba 277-8581, Japan}
\affiliation{Trans-scale Quantum Science Institute, The University of Tokyo, Tokyo 113-0033, Japan}

\author{Nikolaos G. Fytas}
\email{nikolaos.fytas@essex.ac.uk}
\affiliation{School of Mathematics, Statistics and Actuarial Science, University of Essex, Colchester CO4 3SQ, United Kingdom}
\affiliation{Section of Condensed Matter Physics, Department of Physics, National and Kapodistrian University of Athens, Panepistimiopolis, 157 72, Athens, Greece}

\author{Dong-Hee Kim}
\email{dongheekim@gist.ac.kr}
\affiliation{Department of Physics and Photon Science, Gwangju Institute of Science and Technology, Gwangju 61005, South Korea}

\begin{abstract}
Weak first-order transitions can exhibit extended pseudocritical scaling regimes that mimic genuine criticality. We show that the long-debated apparent criticality of the dilute Baxter--Wu model arises from a marginally relevant renormalization-group flow, which ultimately drives the system into a first-order regime with phase coexistence. By integrating conformal perturbation theory with symmetry-resolved tensor-network level spectroscopy, we parameterize a slowly running coupling in a numerically accessible form and construct a generic flow-based diagnostic. The renormalization-group trajectories of the dilute and self-dual Baxter--Wu models under marginal operator deformations collapse onto the multicritical phase diagram of the dilute four-state Potts model, exposing the flow structure on both the marginally relevant and irrelevant sides of the Potts point.
\end{abstract}

\maketitle

\emph{Introduction.}---Weak first-order transitions are ubiquitous in statistical, condensed-matter,
and high-energy physics
\cite{Binder1981, PottsRMP, DeGennes1993, Senthil2024, Bonati2025},
yet diagnosing them unambiguously remains a central challenge
\cite{Nienhuis1979, NauenbergScalapino1980, Cardy1980, Kaplan2009, Wang2017,
GorbenkoRychkovZan2018a, GorbenkoRychkovZan2018b, Kuklov2004, Jiang2008,
Kuklov2008, Harada2013, Chen2013, Nahum2015, Zhao2020, Ma2020, Nahum2020,
Nahum2022, Caci2023, DEmidio2023, Deng2024, Song2025, Yang2026, Flynn2026,
Aharony2022, Hasenbusch2024, Kalz2011, Jin2012, Chatelain2025, Iino2019,
Weber2023, Weber2025, Ma2019, Haldar2023, Jacobsen2024, Tang2024, Kumar2026,
Linden2026}.
At a weak first-order transition, the correlation length remains finite but
can be extremely large, often far exceeding numerically accessible lattice sizes. The result is an extended regime of approximate scale invariance, in which effective exponents appear nearly constant or drift only slowly over many length scales. Such pseudocritical behavior can closely mimic genuine criticality in phenomenological finite-size scaling analyses of conventional observables.  

Renormalization-group (RG) flows provide a natural framework for describing
weak first-order transitions and pseudocriticality.
The two-dimensional $Q$-state Potts model
\cite{Nienhuis1979, NauenbergScalapino1980, Cardy1980}
is a paradigmatic example: for $Q>4$, a ``walking'' RG flow associated with
complex conformal field theories (CFTs)
\cite{Kaplan2009, Wang2017, GorbenkoRychkovZan2018a, GorbenkoRychkovZan2018b}
characterizes the weak first-order transition,
with a correlation length that grows as $\exp{(\pi^2/\sqrt{Q-4})}$
\cite{BuffenoirWallon1993}.
At $Q=4$, the unstable and stable fixed points merge into a multicritical point
governed by a real CFT \cite{NauenbergScalapino1980, Cardy1980},
which connects the second- and first-order transition lines
in the phase diagram of the dilute four-state Potts model
\cite{Cardy1980, Qian2005}.
On the first-order side, the RG flow runs away from the multicritical point,
but extremely slowly: both the correlation length and the latent heat exhibit
an essential singularity as the multicritical point is approached along the
transition line \cite{Cardy1980}.
A quantitative characterization of such slow flows would therefore
provide a sharp diagnostic of weak first-order transitions.

Flow-based diagnostics require direct access to RG trajectories,
which conventional simulations of microscopic models rarely provide.
We focus on diagnosing the flows near the four-state Potts fixed point
under marginal perturbations \cite{NauenbergScalapino1980, Cardy1980}.
Transfer-matrix calculations can extract the marginal scaling field from
the logarithmic finite-size drift of the scaling dimensions,
as demonstrated for the self-dual generalized Baxter--Wu (BW) model
\cite{DengEtAl2010, BloteGuoNightingale2017}.
However, this approach relies on nonlinear multiparameter fits
to separate out irrelevant contributions, which remain significant
at the system sizes accessible to transfer-matrix calculations.
Here we introduce a tensor-network renormalization approach
that enables a parameter-free diagnosis of weak first-order transitions
in BW models.

The critical pure BW model lies at the four-state Potts fixed point
\cite{Baxter1973, Baxter1974, Domany1978, denNijs1979, AlcarazXavier1997,
AlcarazXavier1999, Kinzel1981}.
Upon adding marginal perturbations, it is therefore expected to share
the RG flows of the dilute four-state Potts model \cite{Cardy1980}.
Within this picture, transfer-matrix results for the self-dual
generalized BW model \cite{BloteGuoNightingale2017},
in which the marginal perturbation is an asymmetry between the couplings of
up- and down-pointing triangles,
suggest that a nonzero coupling asymmetry drives
a first-order transition.

Within the same picture, introducing annealed vacancies via a crystal field
should likewise drive the system into the first-order regime.
The seminal study of the dilute BW model \cite{Kinzel1981}
conjectured such an RG trajectory and argued that continuous transitions
are unlikely for any nonzero vacancy fugacity.
Despite this theoretical expectation, the phase diagram of the dilute
BW model remains unsettled.
Subsequent numerical studies
\cite{Costa2004a, Dias2017, Jorge2021, Vasilopoulos2022, Macedo2023, Macedo2024}
support a phase diagram with both second- and first-order transition
lines, although the location of the multicritical point is debated
\cite{Dias2017, Jorge2021, Mataragkas2026}.
As the crystal field increases, the effective exponents drift away from
their Potts values, whereas the central charge remains close
to $c=1$ \cite{Costa2004a, Dias2017, Mataragkas2026}.
These observations may reflect pseudocriticality---an extremely slow but
ultimately runaway RG flow---but this scenario has not been established for
the dilute BW model.

In this Letter, we construct a numerically tractable parameterization of
the running marginal coupling by combining conformal perturbation
theory with symmetry-resolved level spectroscopy.
Using a step-scaling analysis based on tensor-network renormalization,
we show that the dilute BW model follows an extremely slow yet
ultimately runaway RG flow into the first-order regime
for any nonzero vacancy fugacity, along the same trajectory
as the self-dual model with a nonzero coupling asymmetry.
We then map both models quantitatively onto the
multicritical phase diagram of the dilute four-state Potts model,
revealing the flow on both the marginally relevant and the marginally
irrelevant sides of the Potts point.
This slow runaway flow produces an extremely weak drift of the central
charge and a correlation length with an essential singularity that
grows far faster than in the dilute Potts case,
which explains why conventional analyses fail to capture
the extended pseudocritical regime. 

\emph{Models and method.}---We consider two variants of the BW model with marginal perturbations, whose Hamiltonian on the triangular lattice can be written as
\begin{equation} \label{eq:H}
    \mathcal{H} = -\sum_{\boldsymbol{\alpha} = \triangle, \triangledown} J_{\boldsymbol{\alpha}} \sum_{\langle ijk \rangle \in \boldsymbol{\alpha}}
    \sigma_i \sigma_j \sigma_k  - \Delta \sum_i (1 - \sigma_i^2) \,.
\end{equation}
Here $\sigma_i \in \{-1, 0, 1\}$ denotes the spin at site $i$,
and $\Delta$ is the crystal-field strength. 
The sums run over all up-pointing ($\triangle$) and down-pointing ($\triangledown$) triangles,
with three-spin couplings $J_\triangle$ and $J_\triangledown$. 
The first is the dilute BW model
\cite{Kinzel1981, Costa2004a, Dias2017, Jorge2021, Vasilopoulos2022, Macedo2023, Macedo2024, Mataragkas2026},
a spin-$1$ model with symmetric ferromagnetic couplings,
$J_\triangle = J_\triangledown \equiv J > 0$,
where annealed vacancies ($\sigma_i = 0$) are controlled 
by the fugacity $z = \exp(\Delta/T)$ at temperature $T$.
The pure BW model \cite{Baxter1973, Baxter1974} 
is recovered in the limit of $\Delta \to -\infty$.
The second is a self-dual generalization of the pure model
\cite{DengEtAl2010, BloteGuoNightingale2017},
dubbed a ``self-dual'' model hereafter, which is 
a spin-$1/2$ model with asymmetric couplings 
$J_{\triangle, \triangledown} = J \pm \delta J$,
where $\delta J$ is real ($\delta J^2>0$) or purely imaginary ($\delta J^2<0$).

We follow the RG trajectories of both models with loop-optimized
tensor-network renormalization (Loop-TNR)~\cite{Yang2017}.
The partition function of both models is written exactly as a tensor network
of plaquette and copy tensors~\cite{SM}.
We initialize Loop-TNR by coarse-graining each $3\times 3$ block of sites
into a single local tensor. This block matches the period-three pattern of the
spin-reversal symmetries of Eq.~\eqref{eq:H} discussed below, so that every bond
carries a conserved symmetry charge.
The network returns to the same geometry after four coarse-graining steps,
during which the length scale grows by a factor of four.
For level spectroscopy, we extract scaling dimensions from ratios of
transfer-operator eigenvalues
\cite{OkamotoNomura1992, Nomura1995, UedaOshikawa2021}.
Our Loop-TNR calculations use \texttt{TNRKit}~\cite{TNRKit}
with bond dimensions $\chi=48$ and $64$.
For the dilute model, we estimate the transition temperature $T_{\rm c}$
by locating a crossing or jump in partition-function
ratios~\cite{Morita2026}. These $T_{\rm c}$ estimates are consistent
with recent transfer-matrix results~\cite{Mataragkas2026, SM}.
For the self-dual model, $T_{\rm c}$ is fixed
by the self-dual condition, $\sinh(2J/T_{\rm c})=\cosh(2\delta J/T_{\rm c})$.

\emph{Symmetry, CFT spectra, and RG flows.}---Both models share the $S_4$ symmetry of the four-state Potts model. 
The three-spin interactions in Eq.~\eqref{eq:H} are invariant 
under transformations that reverse an even number of spins in every triangle. These transformations generate the Klein four group $V_4$, whose action on the lattice sites partitions them into three sublattices; each nontrivial element reverses the spins on two of the three sublattices. 
The charge carried by each bond of the tensor network is the corresponding $V_4$ charge.
Together with the symmetric group $S_3$ of sublattice permutations, 
the full symmetry group is $V_4\rtimes S_3\simeq S_4$. 
This symmetry is preserved under uniform scalar perturbations, 
such as a crystal field $\Delta$ in the dilute model and a nonzero
$\delta J$ in the self-dual model.
See Supplemental Material (SM)~\cite{SM} for details. 

\begin{figure}[t]
\includegraphics[width=0.48\textwidth]{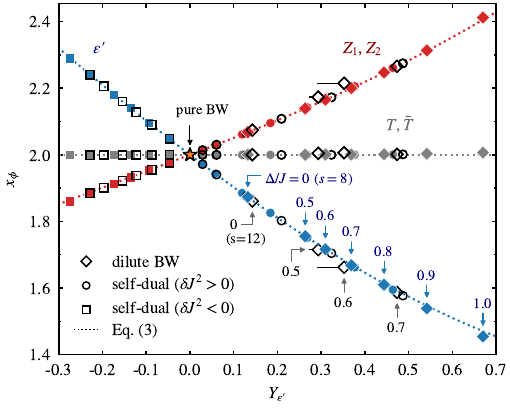}
\caption{Scaling dimensions of the operators $(\varepsilon',T,\bar{T},Z_1,Z_2)$ as
functions of the running coupling $\Yeps$.
Points $(\Yeps,x_\phi)$ are Loop-TNR estimates at coarse-graining steps
$s=8$ (filled) and $s=12$ (open) for the dilute model (diamonds) and the
self-dual model with real (circles) or purely imaginary (squares) $\delta J$. Dotted curves show the perturbed-CFT prediction from Eq.~\eqref{eq:CFTspectrum}.}
\label{fig:splitting}
\end{figure}

At the four-state Potts fixed point, the CFT spectrum contains a fivefold
degeneracy among $\phi \in \{\varepsilon',T,\bar{T},Z_1,Z_2\}$, with scaling
dimension $x_\phi = 2$, where the subleading energy operator $\varepsilon'$ and
the doublet $(Z_1,Z_2)$ are scalar primaries, and $(T,\bar{T})$ are components
of the stress tensor
\cite{Yang1987, DiFrancesco1987, GorbenkoRychkovZan2018b}.
Note that $\varepsilon'$ is the only singlet that generates an
$S_4$-invariant perturbation of the Hamiltonian in Eq.~\eqref{eq:H}.
The coupling to the leading energy operator $\varepsilon$ is tuned to zero at
the transition, and all other relevant operators carry $V_4$ charge.
Perturbation calculations~\cite{SM} show that the marginal coupling to
$\varepsilon'$ shifts $x_{\varepsilon'}$ and $(x_{Z_1},x_{Z_2})$ in opposite
directions, while $(x_{T},x_{\bar{T}})$ remain unchanged.  

The deformation of these marginal operators is closely tied to the RG flow,
and thus to the nature of the transition.
For instance, the shift of the doublet suggests that the drifting exponents
observed in the dilute model are unlikely to originate from a continuous line of
fixed points, as in the Ashkin--Teller model, where an exactly marginal
deformation is generated by the doublet.
The $S_4$-invariant perturbations are not exactly marginal, as indicated by a
nonzero operator product expansion coefficient $C_{\varepsilon'\varepsilon'\varepsilon'}$, which generates 
the leading quadratic term in the RG equation. Derivations are detailed 
in SM~\cite{SM}.

To diagnose an RG trajectory, we introduce a numerically accessible
running coupling that quantifies the strength of the $\varepsilon'$ 
perturbation at length scale $L$.
Near the Potts fixed point, the scaling dimensions admit an expansion in
the renormalized coupling $Y$ of the unit-normalized perturbation
\cite{Cardy1980, GorbenkoRychkovZan2018b, SM}.
Inverting the thermal scaling dimension $x_\varepsilon \approx 1/2 - (3/8)Y$ to leading order, we parameterize the running coupling as
\begin{equation} \label{eq:Ydef}
\Yeps(L)=-\frac{8}{3}\left[x_\varepsilon(L)-\frac12\right].
\end{equation}
Accordingly, the deformation of the marginal operators can be expressed
perturbatively in terms of $\Yeps$ as
\begin{eqnarray} \label{eq:CFTspectrum}
 x_\phi-2 &=& \left(-1,\,0,\,0,\,\tfrac12,\,\tfrac12\right) \Yeps
 +\left(\tfrac14,\,0,\,0,\,\tfrac1{16},\,\tfrac1{16}\right)\Yeps^{2} \nonumber \\ 
 && + \left(\tfrac{3}{32},\,0,\,0,\,\tfrac{9}{64},\,\tfrac{9}{64}\right)\Yeps^{3}
 +O(\Yeps^{4})\, ,
\end{eqnarray}
where $\phi$ is ordered as $(\varepsilon',T,\bar{T},Z_1,Z_2)$.
This expression is independent of microscopic details and thus indicates
whether the two models lie on the same RG trajectory.

The RG flow of the running coupling $\Yeps$ is controlled by the beta function
\cite{Cardy1980,NauenbergScalapino1980,GorbenkoRychkovZan2018b},
\begin{equation} \label{eq:RG}
    \beta(\Yeps) \equiv \frac{\dd \Yeps}{\dd\ln L}
    = \frac{\Yeps^{2}}{2} - \frac{\Yeps^{3}}{8}
    -\frac{11}{128}\,\Yeps^{4} + O(\Yeps^{5})\,,
\end{equation}
written here to three-loop order \cite{SM}.
The leading quadratic term fixes the flow direction: for $\Yeps<0$ the flow approaches the four-state Potts fixed point at $\Yeps=0$ but does not cross it.
For $\Yeps>0$, the flow is very slow near $\Yeps = 0$ but ultimately runs away toward a first-order regime, as we will discuss later.
This behavior is consistent with the perturbed-CFT spectrum in Eq.~\eqref{eq:CFTspectrum}:
a runaway flow corresponds to 
a marginally relevant perturbation with $x_{\varepsilon'}<2$, 
while a continuous transition occurs on the marginally 
irrelevant side with $x_{\varepsilon'}>2$.

\emph{Scale-resolved spectra.}---Figure~\ref{fig:splitting} shows
Loop-TNR estimates of $(\Yeps, x_\phi)$ along the transition line.
Data for both models collapse onto common curves, 
in excellent agreement with the perturbed-CFT prediction in Eq.~\eqref{eq:CFTspectrum}.
We examine crystal fields $\Delta \in [0,1]$ for the dilute model 
and asymmetric couplings $\delta J/T_{\rm c} \in [0,0.23]$ and 
$-i\delta J/T_{\rm c} \in [0,0.26]$ for the self-dual model \cite{SM}. 
The four-state Potts fixed point at $(\Yeps, x_\phi)=(0,2)$ corresponds 
to the pure model.
The dilute model and the self-dual model with real $\delta J$
lie on the $\Yeps>0$ side, implying the same character of the phase transition.
By contrast, the model with purely imaginary $\delta J$ lies 
on the $\Yeps<0$ side.

The RG flow direction can be inferred by comparing spectra at two length scales 
along the tensor-network coarse graining. As shown in Fig.~\ref{fig:splitting}, 
we compare scaling dimensions extracted after $s=8$ and $12$ real-space RG steps, 
which differ by one full period so that the lattice geometry is the same 
but at a larger length scale. The data support the scenario from 
Eqs.~\eqref{eq:CFTspectrum} and \eqref{eq:RG}: on the $\Yeps>0$ side, 
$\Yeps$ increases between these two steps, indicating flow away from the fixed point.
This trend holds for all field values examined in the dilute model 
and in the self-dual model with real $\delta J$.

By contrast, $\Yeps<0$ occurs only for purely imaginary $\delta J$ 
in the self-dual model.
There, $\Yeps$ increases (i.e., $|\Yeps|$ decreases) between $s=8$ and $12$, 
indicating flow toward the fixed point at $\Yeps=0$.
These step-to-step changes provide only local snapshots of the flow 
for two specific RG steps and a limited parameter range.
Below we show that, nevertheless, they collapse onto a unified step-scaling 
picture consistent with the beta function.

\begin{figure}[t]
\includegraphics[width=0.48\textwidth]{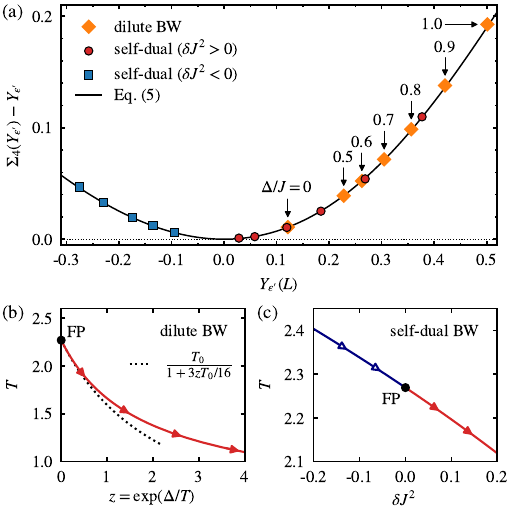}
\caption{ Step scaling of the running coupling $\Yeps$.
(a) Loop-TNR estimates of the increment $\Yeps(s+4)-\Yeps(s)$ for the dilute model (diamonds, $s=6$) and the self-dual model (circles and squares, $s=8$).
The solid line shows the theoretical increment $\Sigma_4(\Yeps)-\Yeps$,
obtained by solving Eq.~\eqref{eq:ssf} for each value of $\Yeps$.
For the dilute model, the value of $\Delta$ is annotated next to each diamond. The corresponding RG flows along the transition line are shown in the phase
diagrams of the dilute (b) and self-dual (c) models.
Filled (open) arrows indicate first-order (second-order) transitions.}
\label{fig:stepscaling}
\end{figure}

\emph{Step-scaling analysis.}---Equation~\eqref{eq:RG} defines 
the beta function for an infinitesimal change of length scale.
However, real-space renormalization in lattice systems proceeds 
with discrete coarse-graining steps, leading to a map between running couplings.
To describe the flow of $\Yeps$ in our tensor-network RG scheme, 
we use step-scaling analysis \cite{LuscherWeiszWolff1991} and 
define the map $\Yeps \mapsto \Sigma_4(\Yeps)$, where $\Sigma_4(\Yeps)$ 
is the running coupling after four coarse-graining steps.
For a given $\Yeps$, the function $\Sigma_4(\Yeps)$ is determined by solving
\begin{equation} \label{eq:ssf}
    \int_{\Yeps}^{\Sigma_4(\Yeps)} \frac{1}{\beta(Y)}\,\dd Y = \ln 4\,.
\end{equation}
The increment $\Sigma_4(\Yeps)-\Yeps$ provides a step-scaling analogue of 
the beta function, as follows from the leading-order quadratic dependence 
$\Sigma_4(\Yeps)-\Yeps \approx \Yeps^2 \ln 2$.
This increment enables direct comparison with the step-to-step change 
measured in the microscopic models.

Figure~\ref{fig:stepscaling} shows that the Loop-TNR data collapse 
onto the theoretical step-scaling curve obtained from 
the three-loop beta function. 
The resulting non-negative, U-shaped curve encodes the direction (sign) 
and rate (magnitude) of the RG flow on either side of the fixed point,
providing a unified framework for characterizing
the phase transitions in both models.
In the self-dual model with purely imaginary $\delta J$
($\Yeps < 0$), the flow rate vanishes
as it approaches the fixed point at $\Yeps = 0$, manifesting 
as strong logarithmic corrections \cite{SalasSokal1997}.
In the dilute model and in the self-dual model with real $\delta J$
($\Yeps > 0$), the flow instead runs away but remains extremely slow 
near the fixed point, leading to an essential singularity in 
the correlation length and hence pseudocriticality.

\begin{figure}[t]
\includegraphics[width=0.48\textwidth]{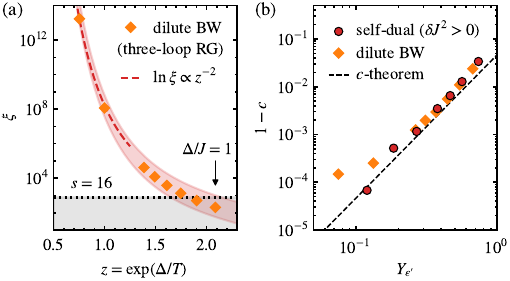}
\caption{Essential singularity and pseudocriticality.
(a) Correlation length $\xi$ along the transition line of the dilute model
as a function of the vacancy fugacity $z$.
Diamonds show the three-loop beta-function result with
$Y_{\varepsilon',0}=\Yeps(s=8)$ and $Y_\mathrm{IR}=1$.
The shaded region is bounded by the curves with $Y_\mathrm{IR}=0.7$ and $1.5$.
For comparison, the length scale corresponding to $s=16$, which is needed to
resolve phase coexistence at $\Delta/J=1$, is also indicated.
(b) Central-charge drift $1-c$ as a function of $\Yeps$, measured at $s=8$.
The dashed line shows the $c$-theorem prediction,
$c \approx 1-(3/64)\,\Yeps^3$.}
\label{fig:pseudocriticality}
\end{figure}

\emph{Essential singularity and pseudocriticality.}---To understand the implications of the slow RG flow, one needs to connect
the microscopic model parameters to the running coupling $Y_{\varepsilon',0}\equiv Y_{\varepsilon'}(L_0)$ at a reference length scale $L_0$
beyond which finite-size corrections from irrelevant operators are negligible.
In the self-dual model with $\delta J\neq 0$, invariance under 
$\delta J\to-\delta J$ implies $Y_{\varepsilon',0} \propto \delta J^2$ 
to leading order \cite{BloteGuoNightingale2017}.
For the dilute model with $N$ sites, the grand partition function 
at inverse temperature $K = J/T$ and fugacity $z = e^{\Delta/T}$ 
can be written as
\begin{equation}
    Z_G(K,z) = e^{5\sqrt{2}Nz/16}\, Z_\mathrm{BW}(\tilde{K})
             + \sum_{N_v = 2}^N z^{N_v} Z_{N_v}(K)\,,
\end{equation}
where $N_v$ is the number of vacancies and $Z_{N_v}(K)$ is 
the partition function at fixed $N_v$
(see SM~\cite{SM} for details). The single-vacancy contributions 
are absorbed into the partition function
$Z_\mathrm{BW}(\tilde{K})$ of the pure model with $\tilde{K}=K-3z/16$. 
Thus, a single vacancy shifts the transition temperature as
$T_{\rm c}(z) \approx T_0/(1+3zT_0/16)$ for $z\ll 1$, where
$T_0=2/\ln(1+\sqrt{2})$, but does not generate an RG flow.
The leading contribution in $Y_{\varepsilon',0}$ then arises at order $z^2$. 

For $\Yeps>0$, integrating the one-loop beta function 
$\beta(\Yeps)=\Yeps^2/2$ from $Y_{\varepsilon',0}$ up to 
an $O(1)$ infrared scale relates the correlation length $\xi$ 
to $Y_{\varepsilon',0}$ as 
$\xi/L_0\propto\exp(2/Y_{\varepsilon',0})$~\cite{SM}.
The resulting essential singularity is therefore governed 
by the parameter dependence of $Y_{\varepsilon',0}$.
Figure~\ref{fig:pseudocriticality}(a) plots $\xi$ obtained numerically 
from the three-loop beta function in Eq.~\eqref{eq:RG}, 
using $L_0=48$ ($s=8$) and infrared values 
$Y_\mathrm{IR} \in [0.7, 1.5]$ in the dilute model.
For small $z$, the singularity appears as $\ln \xi \sim z^{-2}$. 
Using $T_0-T_{\rm c}(z)\propto z$, this can be rewritten as 
$\xi\sim\exp[a(T_0-T_{\rm c})^{-2}]$ with $a>0$,
which grows rapidly as $T_{\rm c}(z)$ approaches $T_0$ 
along the transition line.
Although the exponential growth of $\xi(T)$ closely 
resembles the first-order behavior found in the dilute four-state Potts model, 
the singularity with $\ln\xi \propto (T_0-T_{\rm c})^{-2}$ 
in the dilute BW model is much stronger than 
in the Potts counterpart, where
$\ln\xi \propto (T_0-T_{\rm c})^{-1}$~\cite{Cardy1980}.   

As $z$ increases (and $\xi$ decreases) 
along the transition line, the drift of the effective exponents becomes more apparent.
In particular, the thermal scaling field $\varphi_t$ obeys an RG equation,
\begin{equation}
    \frac{\dd \varphi_t}{\dd \ln L} = (2-x_\varepsilon) \varphi_t
    = \frac{3}{8}\big[4 + \Yeps(L)\big]\varphi_t \, .
\end{equation}
In the weak-coupling regime, the flow slightly away from the transition line 
is dominated by $\varphi_t \sim L^{y_t}$, where 
$y_t \approx 3/2 + 3Y_{\varepsilon',0}/8$, neglecting higher-order 
logarithmic terms in $\Yeps(L)$.
This accounts for the apparent finite-size scaling and
the drift of the thermal exponent at a larger $\Delta$, 
as observed in previous Monte Carlo calculations
\cite{Jorge2021, Vasilopoulos2022, Macedo2023, Mataragkas2026}.

The drift of the central charge $c$ provides further evidence of
pseudocriticality. The $c$-theorem~\cite{Zam1986} states that 
the central charge decreases monotonically along the runaway RG flow
for $\Yeps>0$. Solving the $c$-function equation,
$\frac{\dd c}{\dd \ln L} = - \frac{9}{32}\beta(Y)^2$,
rewritten for our definition of the running coupling, gives
$c(\Yeps) \approx 1 - \frac{3}{64}\Yeps^3$ \cite{SM}.
This $O(\Yeps^3)$ drift is far more difficult to resolve numerically
than the $O(\Yeps)$ drift in the effective exponent. This numerical subtlety may
underlie previous transfer-matrix reports of a persistent $c\approx 1$ over a
wide range of $\Delta$ \cite{Costa2004a, Dias2017, Mataragkas2026}.
Remarkably, our Loop-TNR calculations resolve this systematic drift in $c$,
in excellent agreement with the $c$-theorem prediction
as presented in Fig.~\ref{fig:pseudocriticality}(b).

\begin{figure}[t]
\includegraphics[width=0.48\textwidth]{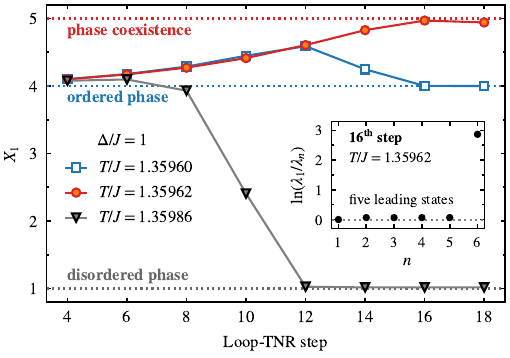}
\caption{Partition-function ratio $X_1$ for the dilute model at $\Delta/J=1$. The ordered and disordered phases flow to $X_1=4$ and $X_1=1$, respectively, while $X_1=5$ marks the first-order transition at
$T_{\rm c}/J \approx 1.35962$. The inset shows the five nearly degenerate largest eigenvalues $\lambda_{1},\dots,\lambda_{5}$ of the transfer-operator spectrum at coarse-graining step $s=16$, indicating five-phase coexistence at the
transition.} 
\label{fig:coexistence}
\end{figure}

\emph{Phase coexistence.}---Despite the evidence for a weak first-order transition 
discussed above, the exponential growth of $\xi$ makes direct first-order signatures 
elusive in the weak-coupling regime. Away from weak coupling, recent multicanonical 
Monte Carlo simulations reported a finite surface tension at $T=1.1$ ($z \approx 4$) \cite{Mataragkas2026}.
We connect these regimes by identifying a spectral signature of 
phase coexistence at $\Delta/J=1$ ($z \approx 2$) in our Loop-TNR calculations. 

We use the partition-function ratio $X_1=\sqrt{X_hX_v}$~\cite{Morita2026},
where $X_{h,v}=(\operatorname{Tr}M_{h,v})^2/\,\operatorname{Tr}M_{h,v}^2$ and
$M_h$ ($M_v$) is the horizontal (vertical) transfer operator.
In the thermodynamic limit, a first-order transition is signaled by
a discontinuous jump in $X_1$. In the dilute model, five-phase coexistence
corresponds to $X_1=5$, reflecting a fivefold degeneracy of the leading
transfer-operator eigenvalues; the ordered and disordered phases similarly give
$X_1=4$ and $1$, respectively.
At finite length scales, the spectral gaps remain nonzero, leading to avoided
crossings near the transition. Nonetheless, as shown in Fig.~\ref{fig:coexistence},
the discrete values of $X_1$ are fully resolved at Loop-TNR step $s=16$,
locating phase coexistence at $T_{\rm c}/J \approx 1.35962$.
This also indicates that the length scale $L=768$ at $s=16$ exceeds 
$\xi$ for $\Delta/J=1$, corroborating the estimates 
in Fig.~\ref{fig:pseudocriticality}(a).
The phase-coexistence spectrum found in the RG flow at $\Delta/J=1$ 
extends our step-scaling analysis beyond the upper end of 
the weak-coupling regime and confirms that the flow ultimately 
runs away toward a first-order transition.

\emph{Summary and outlook.}---We have introduced a tensor-network diagnostic for slow RG flows
induced by a marginal perturbation of the four-state Potts fixed point, and
applied it to the dilute and self-dual BW models.
Using Loop-TNR to perform level spectroscopy and step-scaling analysis,
we quantitatively verified the perturbed-CFT predictions and the
corresponding three-loop RG equation.
This enables a unified mapping of both models onto the RG flow of the
four-state Potts theory: a marginal fixed point separates a second-order
regime with marginally irrelevant flow from a first-order regime with
marginally relevant flow.
In particular, for any finite vacancy fugacity the dilute BW model
lies on the marginally relevant side, indicating a first-order transition
throughout the phase diagram.

Our results have two main implications.
First, they resolve the long-standing debate on the phase diagram of the
dilute BW model: the runaway RG flow rules out a true second-order
transition, while the extremely slow flow near the fixed point produces
pseudocritical behavior at accessible length scales.
Second, our approach extends tensor-network renormalization to a quantitative probe of RG flows at weak first-order transitions.
The protocol involves no fitting parameters and provides direct numerical
access to the RG flow equation, as demonstrated here for the four-state
Potts case.

Promising generalizations include lattice realizations of complex CFTs
\cite{Ma2019, Haldar2023, Weber2023, Weber2025, Jacobsen2024, Tang2024,
Linden2026, Kumar2026}, where walking RG flows are expected
\cite{GorbenkoRychkovZan2018a, GorbenkoRychkovZan2018b}.
While the present study establishes a diagnostic for slow flows
near a fixed point described by a real CFT, extending tensor-network
renormalization to complex coupling space remains an important open question.
Another direction is to generalize our diagnostic to a broader class
of fixed points beyond weak first-order transitions, for instance
by tuning marginal perturbations to enlarge effective symmetries.

\emph{Acknowledgments.}---H.-Y.~L. was supported by the Basic Science Research Program through the National Research Foundation (NRF) of Korea funded by the Ministry of Science and ICT (MSIT) (Grant Nos. RS-2023-00220471 and RS-2025-16064392). H.-Y.~L also acknowledges the hospitality of the Institute for Solid State Physics, The University of Tokyo, where part of this work was carried out. N.~K. acknowledges support from JSPS KAKENHI Grant No.~23K25789. The work of N.G.~F. was supported by the Leverhulme Trust through a Research Project Grant (RPG-2026-025). D.-H.~K. acknowledges support from the NRF of Korea (Grant Nos. RS-2024-00392445, RS-2025-02306096, RS-2025-25464697) funded by MSIT.

\emph{Data availability.}---The data that support the findings of this article are not publicly available. The data are available from the authors upon reasonable request.

\bibliography{refs}

\end{document}